\input harvmac 
\input epsf.tex
\overfullrule=0mm
\newcount\figno
\figno=0
\def\fig#1#2#3{
\par\begingroup\parindent=0pt\leftskip=1cm\rightskip=1cm\parindent=0pt
\baselineskip=11pt
\global\advance\figno by 1
\midinsert
\epsfxsize=#3
\centerline{\epsfbox{#2}}
\vskip 12pt
{\bf Fig. \the\figno:} #1\par
\endinsert\endgroup\par
}
\def\figlabel#1{\xdef#1{\the\figno}}
\def\encadremath#1{\vbox{\hrule\hbox{\vrule\kern8pt\vbox{\kern8pt
\hbox{$\displaystyle #1$}\kern8pt}
\kern8pt\vrule}\hrule}}

\def\tvi{\vrule height 12pt depth 6pt width 0pt}
\def\tv{\tvi\vrule}

\Title{T94/073}
{{\vbox {
\bigskip
\centerline{Folding Transition of the Triangular Lattice}
}}}
\bigskip
\centerline{P. Di Francesco}
\medskip
\centerline{and} 
\medskip
\centerline{E. Guitter,}

\bigskip

\centerline{ \it Service de Physique Th\'eorique de Saclay
\footnote*{Laboratoire de la Direction des Sciences 
de la Mati\`ere du Commissariat \`a l'Energie Atomique.},}
\centerline{ \it F-91191 Gif sur Yvette Cedex, France}

\vskip .5in

We study the problem of folding of the regular triangular lattice in
the presence of bending rigidity $K$ and magnetic field $h$
(conjugate to
the local normal vectors to the triangles).  A numerical
study of the transfer matrix of the problem shows the existence of 
three first order transition lines in the $(K,h)$ plane separating three
phases: a folded phase, a phase frozen in the completely flat 
configuration (with all normal vectors pointing up) and its mirror image
(all normal vectors pointing down).
At zero magnetic field, a first order folding transition is found at a 
positive value $K_c\simeq 0.11(1)$ of the bending rigidity, corresponding
to a triple point in the phase diagram.

\noindent
\Date{06/94}


\nref\NP{D.R. Nelson and L. Peliti, J. Phys. France {\bf 48} (1987) 1085.}
\nref\KN{Y. Kantor and D.R. Nelson, Phys. Rev. Lett. {\bf 58} (1987) 2774
and Phys. Rev.  {\bf A 36} (1987) 4020.}
\nref\PKN{M. Paczuski, M. Kardar and D.R. Nelson, Phys. Rev. Lett. {\bf 60} (1988) 2638.}
\nref\DG{F. David and E. Guitter, Europhys. Lett. {\bf 5} (1988) 709.}
\nref\TAS{see also 
M. Baig, D. Espriu and J. Wheater, Nucl. Phys. {\bf B314} (1989) 587;
R. Renken and J. Kogut, Nucl. Phys. {\bf B342} (1990) 753; R. Harnish and
J. Wheater, Nucl. Phys. {\bf B350} (1991) 861.}
\nref\KJ{Y. Kantor and M.V. Jari\'c, Europhys. Lett. {\bf 11} (1990) 157.}
\nref\DIG{P. Di Francesco and E. Guitter, Europhys. Lett. {\bf 26} (1994)
455.}
\nref\BT{R.J. Baxter and S.K. Tsang, J. Phys. {\bf A13} Math. Gen. (1980) 1023.}
\nref\BAX{R.J. Baxter, J. Math. Phys. {\bf 11} (1970) 784 and 
J. Phys. {\bf A19} Math. Gen. (1986) 2821.}


\newsec{Introduction}

It is tempting to try to describe geometrical objects like 
polymers (1D) or membranes (2D) in analogy with spin systems.
Natural spin variables are provided for instance by the local normal or
tangent vectors to the object, while elastic properties like bending
rigidity naturally translate into some nearest neighbor spin coupling.
However, the correspondence between geometrical objects and spin systems
can be subtle, especially in two dimensions, where geometric
constraints on, say, the normals to the membrane imply local 
constraints on the associated spin variables. 
Such constraints play a crucial role for
tethered  membranes, {\it i.e.} 2D polymerized networks 
with fixed connectivity,
since they induce a crumpling transition by
stabilizing an ordered phase in a region where the unconstrained
spin system would be disordered. This phenomenon was recognized in
[\xref\NP -\xref\TAS], where a continuous crumpling transition
is predicted.

Such a drastic change of statistical behavior is observed in the
present paper, 
where we consider a spin system describing the thermodynamics of
folding of the regular triangular lattice, a problem first 
considered in \KJ.
Considered as a geometrical object,
the lattice describes a tethered membrane skeleton, made of
rigid bonds along which folds can be performed. 
Here we consider only {\it complete} foldings which result in 
{\it two--dimensional}
folded configurations of the membrane.
In such a process, each bond 
serves as a hinge between its two neighboring triangles, and
is in either one of the two states: 
folded (with the two neighboring triangles
face to face) or not (side by side)\foot{We refer the reader to
[\xref\KJ -\xref\DIG] for a more formal definition of folding.}.  
A folding configuration
(folded state) of the system is entirely 
specified by the list of its folded bonds.
This definition corresponds to a ``phantom" membrane, where 
the folding process may imply
self--intersections, and where one 
cannot distinghish in the folded state between 
different piling orders for superimposed triangles.

%
\fig{The eleven local fold environments for a vertex. Folds are represented by
thick lines. One of the two possible spin configurations on the triangles
is also indicated.}
{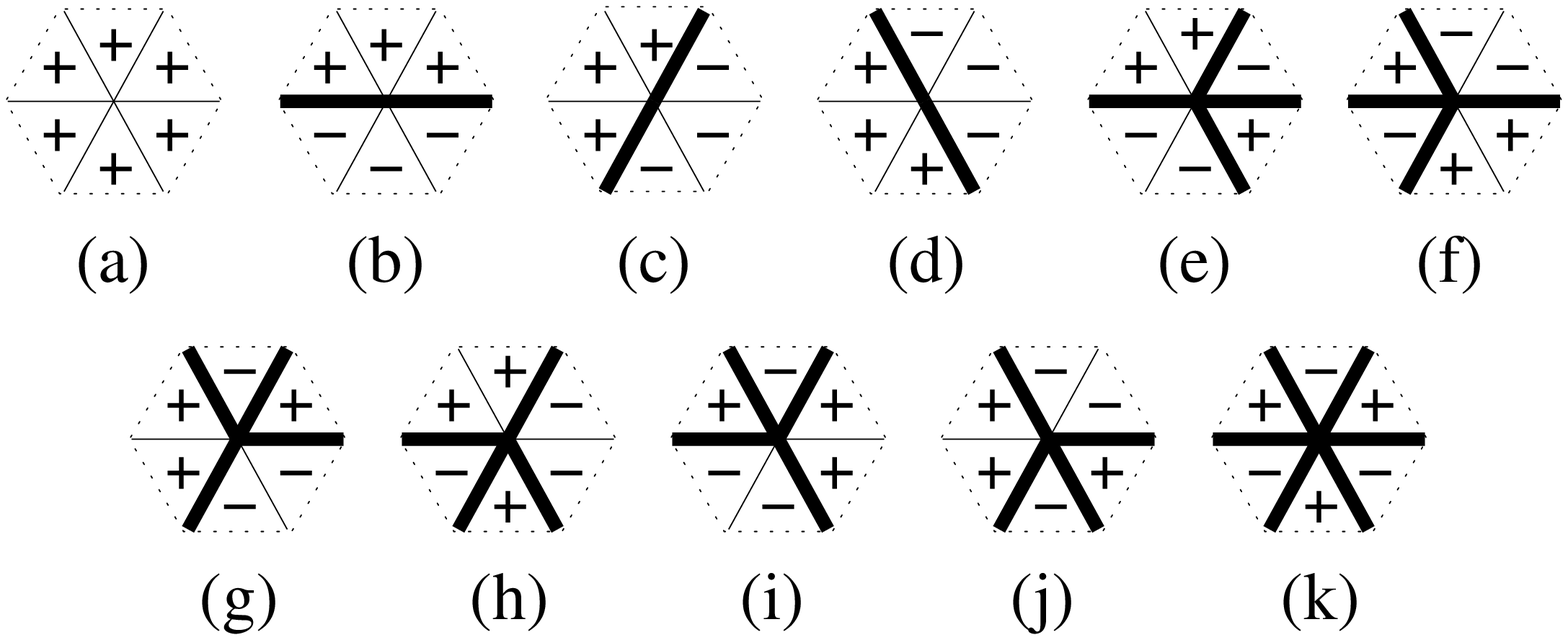}{10. truecm}
\figlabel\eleven
%
With this simplified definition, our folding problem can be formulated
as an eleven vertex model, expressing that 
the immediate surroundings of a vertex in a folded state must be in one 
of the eleven local configurations depicted on
Fig.\eleven. 
In spite of its local definition, 
folding is a highly non--local operation.
As explained in \DIG, 
we note that whenever a bond is folded, say, on the left half of a vertex,
then another bond is folded on the right half, hence folds propagate 
throughout the lattice.

In a previous work \DIG, we have computed the exact thermodynamic 
entropy per triangle which counts the number
$Z_{\rm f}$ of folded configurations for a finite lattice 
made of $N_t$ triangles for large $N_t \to \infty$.
This was done  
by mapping the eleven vertex model above onto the 
three--coloring problem 
of the triangular lattice bonds, 
solved exactly by Baxter \BAX\ in its dual
version, the three--coloring of the hexagonal lattice.
The result reads
\eqn\entro{
\eqalign{s&\equiv \lim_{N_t \to \infty} {1 \over N_t} {\rm Log} \ 
Z_{\rm f}\equiv {\rm Log} \, q \cr
q&=\prod_{p=1}^{\infty} {3p-1 \over \sqrt{3p(3p-2)} }=
{\sqrt{3} \over 2 \pi} \Gamma(1/3)^{3/2} =1.20872...\ . \cr}}

As mentioned above, we will rather use here
the alternative description of the model
in terms of Ising spin variables
$\sigma=\pm 1$ defined on the triangles, and
indicating whether they face up or down in the folded state. 
One can think of the spin as the normal vector to the triangle.
Spin configurations are given together with the fold configurations
on Fig.\eleven. 
Note that there are two spin configurations for each folded state, due
to the degeneracy under reversal of all spins, hence the partition
function $Z$ of the spin system is twice that of the eleven vertex model:
$Z=2 \, Z_{\rm f}$.
It is clear from Fig.\eleven\ that the only allowed vertex 
environments are those with exactly $0$, $3$ or $6$ surrounding up spins. 
In order for a spin configuration to correspond to a folded
state, the six spins $\sigma_i$
around any vertex $v$ must satisfy the {\it local constraint}
\eqn\locons{\Sigma_v \equiv \sum_{i\ {\rm around} \ v} \sigma_i 
\ =\  0 \ {\rm mod} \ 3\ ,}
since $\Sigma_v=2(\hbox{ number of up spins}) -6$ is a multiple of $3$
iff the number of up spins itself is a multiple of $3$.
Beyond the above counting of 
the number of allowed constrained spin configurations, 
it would be desirable to understand the
effect of a bending energy for the folds, 
characterizing the rigidity of the membrane.
In the spin language, this means the presence of a 
ferromagnetic Ising--like 
interaction energy $- J \sigma_i \sigma_j$ between nearest neighbors. 
Most properties of the folded tethered 
membrane can in fact be investigated
by studying the magnetic behavior of our
constrained Ising spin system.
The average magnetization $M$ of the system is indeed an order 
parameter which is characteristic of the flatness of the membrane
($|M|>0$ for a configuration which is flat in average, and $M=0$ for a
configuration which is folded in average).  
This suggests to introduce a magnetic field $H$ in the system
(with energy $-H \sigma_i$ per triangle),
with no direct physical meaning for the membrane, 
but instrumental in revealing information on its average state of folding.
In the following, we will therefore consider the constrained Ising
model with Hamiltonian
\eqn\hamising{ {\cal H}_{\rm Ising} = 
-J \sum_{(ij)} \sigma_i \sigma_j - H \sum_i \sigma_i \ . }
For convenience we will use the
reduced coupling and magnetic field
\eqn\conv{K \equiv {J/k_BT} \ \ ; \ \  h \equiv H/k_BT \ .}
In our study of this model, we will give numerical and 
theoretical evidence for the 
existence of a {\it first order transition line} in the $(K,h)$ plane, 
between an ordered phase $M=1$ (for $h>0$)
where the membrane is completely flat, and a disordered phase $M=0$, where
the membrane is folded and has a non--vanishing entropy.
The most surprising fact is that we find no intermediary magnetization
of the system in the thermodynamic limit.
For $h=0$, a first order folding transition still takes place, at
a critical value $K_c$ of the reduced Ising coupling $K$.
All these results clearly show a drastic modification of the 
thermodynamics of the standard Ising model, emphasizing the special
role played by the constraint \locons.

The paper is organized as follows. In section 2, we describe the transfer 
matrix that we shall use for numerical simulations on the
thermodynamics of the constrained spin system, and show how to take 
advantage of some particular properties of this matrix.
The results for the magnetization in the 
presence of a magnetic field are discussed in section 3,
and lead us to formulate the abovementioned two phase ($M=0,1$) hypothesis. 
Under this
assumption, we also derive a simple argument to calculate 
the critical value
of the magnetic field where the transition between these phases takes place.
Section 4 is dedicated to the precise study of the first order
transition line in the thermodynamic limit.
In particular, we find the critical value $K_c$
beyond which the $M=1$ phase persists even at zero magnetic field.
We discuss the general phase diagram of the system in section 5, and 
gather more evidence for the first order character
of the transition. 
Related topics are discussed in section 6, including the exact 
solution for the square lattice as well as some predictions of a
possible {\it antiferromagnetic transition} within the $M=0$ phase,
for negative $K$. Section 7 is a brief conclusion.

\newsec{Transfer matrix description}

We consider the folding of an infinite strip of
triangular lattice of finite width $L$, with free boundary conditions on 
the edges of the strip. 
Imposing periodic boundary conditions
at infinity, the partition function of the model is expressed as
\eqn\pfunct{Z^{(L)}(K,h)=\lim_{N  \to \infty}  
\big[ {\rm Tr}(T^{(L)}(K,h)^N) \big]^{1 \over N},}
where $T^{(L)}$ denotes the transfer matrix, acting as an operator
on a column of size $L$, whose state is specified by the $2L$ spin values
$\sigma_i=\pm 1, i=1,..,2L$ on the triangles.

%
\fig{The transfer matrix $T^{(L)}$. }{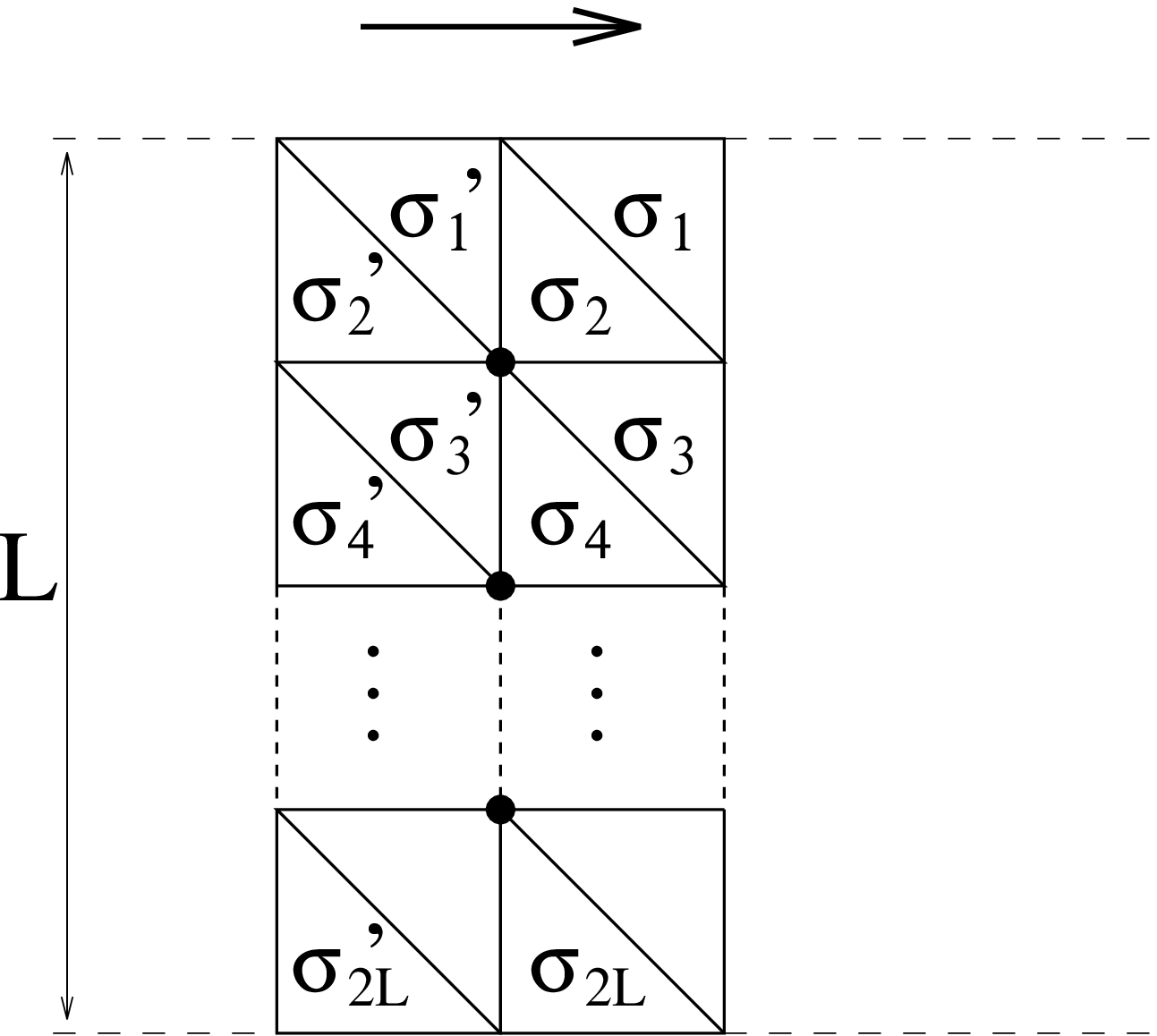}{6cm}
\figlabel\transmat
%
The matrix element
of $T^{(L)}$ between two consecutive columns, as depicted on Fig.\transmat, 
reads
$$T^{(L)}_{\{ {\bf \sigma'} \},\{ {\bf \sigma} \}}(K,h)=
T^{(L)}_{\{ {\bf \sigma'} \},\{ {\bf \sigma} \}}(0,0) \ \
U^{(L)}_{\{ {\bf \sigma'} \},\{ {\bf \sigma} \}}(K) \ \
V^{(L)}_{\{ {\bf \sigma'} \},\{ {\bf \sigma} \}}(h),$$
where 
$T^{(L)}(0,0)$ imposes the local folding constraint \locons\
on the six spins
surrounding each of the $L-1$ inner vertices (marked by black dots
on Fig.\transmat\ ), 
and $U^{(L)}$ and $V^{(L)}$ are the usual
temperature and magnetic field contributions to the transfer
matrix of the Ising model, namely
$$\eqalign{
T^{(L)}_{\{ {\bf \sigma'} \},\{ {\bf \sigma} \}}(0,0)&=
\prod_{i=1}^{L-1} \delta(\sigma_{2i}+\sigma_{2i+1}+\sigma_{2i+2}+
\sigma_{2i-1}'+\sigma_{2i}'+\sigma_{2i+1}' \ {\rm mod} \ 3) \cr
U^{(L)}_{\{ {\bf \sigma'} \},\{ {\bf \sigma} \}}(K)
&=\exp({{K \over 2} \sum_{i=1}^{2L-1} (\sigma_i \sigma_{i+1}
+\sigma_i' \sigma_{i+1}')+ K \sum_{i=1}^L \sigma_{2i} 
\sigma_{2i-1}' }) \cr
V^{(L)}_{\{ {\bf \sigma'} \},\{ {\bf \sigma} \}}(h)&=
\exp({ {h \over 2} \sum_{i=1}^{2L} (\sigma_i +\sigma_i') }) ,\cr}$$
with $\delta(x)$ the usual Kronecker delta function on integers.

In the large $N$ limit, the partition function \pfunct\ is 
dominated by the largest eigenvalue of
$T^{(L)}(K,h)$, which we denote by $\lambda_{\rm max}^{(L)}(K,h)$,
and the corresponding free energy per triangle reads
$$ -F^{(L)}(K,h)={1 \over 2L} {\rm Log} \, Z^{(L)}(K,h) =
{1 \over 2L} {\rm Log} \, \lambda_{\rm max}^{(L)}(K,h) \ .$$
The thermodynamic free energy per triangle is then defined as the 
$L \to \infty$ limit
\eqn\freenrj{ -F(K,h)= \lim_{L \to \infty}\ {1 \over 2L} {\rm Log} 
\, \lambda_{\rm max}^{(L)}(K,h) \ .}
Our main task is therefore the numerical extraction of this eigenvalue,
together with the corresponding eigenvector $v_{\rm max}^{(L)}(K,h)$.
We will also be interested in evaluating the subleading eigenvalue
$\lambda_{\rm sub}^{(L)}(K,h)$.

The analysis can be simplified in view of the following
properties of the transfer matrix. 
Due to the local constraint on the spin
variables, the matrix $T^{(L)}(K,h)$, of size $2^{2L} \times 2^{2L}$,
is sparse. The exact number $N_L$ of non--zero entries of $T^{(L)}$ is
easily computed by using the {\it vertical} transfer matrix
of size $2$, $T^{(2)}(0,0)$, namely
$$N_L = \sum_{\{{\bf \sigma'}\},\{ {\bf \sigma} \}} \bigg[ 
T^{(2)}(0,0)^L \bigg]_{\{{\bf \sigma'}\},\{ {\bf \sigma} \}} .$$
Diagonalizing $T^{(2)}$ exactly, we find
$$N_L = {1 \over 2} \bigg[ \bigg({7 + \sqrt{17} \over 2} \bigg)^L 
\bigg\{ {11+3 \sqrt{17} \over \sqrt{17}}\bigg\} 
-\bigg({7 - \sqrt{17} \over 2}\bigg)^L
\bigg\{ {11-3 \sqrt{17} \over \sqrt{17}}\bigg\}  \bigg] \ . $$
The first values of $N_L$ are given in Table I below.
$N_L$ grows like $(5.56..)^L \ll 16^L$, which justifies the use of 
a diagonalization algorithm adapted to 
sparse matrices: starting from a given vector, we compute recursively
its iterated image by $T$, which we moreover
normalize at each step; the process converges to the eigenvector $v_{\rm max}$
associated to $\lambda_{\rm max}$.
The subleading eigenvalue $\lambda_{\rm sub}$ is obtained by
use of the same algorithm, together with a suitable projection procedure,
guaranteeing at each step that we substract the component along
the maximum eigenvector.
This projection procedure is made easy by the following symmetry property 
of the transfer matrix.
Consider the picture of Fig.\transmat, and rotate it by $180$ degrees.
One gets an identity between two transfer matrix elements, namely
$$ 
T^{(L)}(K,h)_{\sigma_1',\sigma_2',...,\sigma_{2L}';
\sigma_1,\sigma_2,...,\sigma_{2L}} =
T^{(L)}(K,h)_{\sigma_{2L},...,\sigma_2,\sigma_1;
\sigma_{2L}',...,\sigma_2',\sigma_1'} $$
This can be recast into $T^{(L)}(K,h)^T= R\, T^{(L)}(K,h) \, R$, where $R$
is an involution, with matrix elements
$$R_{\sigma_1',\sigma_2',...,\sigma_{2L}';
\sigma_1,\sigma_2,...,\sigma_{2L}} = \prod_{i=1}^{2L} 
\delta(\sigma_i'-\sigma_{2L+1-i}).$$
Any eigenvector $v$ with eigenvalue
$\lambda<\lambda_{\rm max}$ is then clearly
orthogonal to $R\, v_{\rm max}$. To project out
the $v_{\rm max}$ component off a given vector $w$, one simply has
to perform the substitution 
$$w \to w- {( R \, v_{\rm max}|w) \over
( R \, v_{\rm max}|v_{\rm max})} v_{\rm max}.$$

{}From the knowledge of the eigenvector $v_{\rm max}^{(L)}(K,h)$,
we can get the expectation value of {\it local} observables of the
system. For instance, the magnetization $M$ is obtained
as follows.  
The magnetization operator $\mu$ acts diagonally on the columns
of $2L$ spins,
with diagonal elements 
$\mu_{\{ {\bf \sigma}\},\{ {\bf \sigma}\}}=\sum_{i=1}^{2L} \sigma_i$.
In the $N \to \infty$ limit,
its expectation value $M=\langle \mu \rangle$ reads
\eqn\magnet{ M= {(R \, v_{\rm max}^{(L)}(K,h)|\mu \,
v_{\rm max}^{(L)}(K,h)) \over 
(R \, v_{\rm max}^{(L)}(K,h)|v_{\rm max}^{(L)}(K,h) )} \ .}

As a test of the precision of our algorithms, we first compute 
$\lambda_{\rm max}^{(L)}(0,0)$ for various sizes $L=2,3,..,9$. 
The results are summarized in Table I.
 
%
$$\vbox{\offinterlineskip
\halign{\tv\quad # & \quad\tv \quad 
# & \quad \tv \quad  # & \quad \tv #\cr 
\noalign{\hrule}
\tvi $L$ & $N_L$& $\lambda_{\rm max}^{(L)}(0,0)$ &\cr
\noalign{\hrule}
\tvi  $2$ &  $88$ &        $5.56155281$ &\cr
\tvi  $3$ &  $488$ &       $7.89397081$ &\cr
\tvi  $4$ &  $2712$ &     $11.32598261$ &\cr
\tvi  $5$ &  $15080$ &    $16.34742307$ &\cr
\tvi  $6$ &  $83864$ &    $23.67855022$ &\cr
\tvi  $7$ &  $466408$ &   $34.37351897$ &\cr
\tvi  $8$ &  $2593944$ &  $49.97256996$ &\cr
\tvi  $9$ &  $14426344$ & $72.72483625$ &\cr
\noalign{\hrule} }} $$
\centerline{{\bf Table I:} The number $N_L$ of non--vanishing transfer
matrix elements and }
\centerline{the maximum eigenvalue $\lambda_{\rm max}$ for
strips of width $L=2,3,..,9$.}
\vskip 1cm
%
\noindent This leads to an estimate\foot{We obtain the value 
of $q$ as the limit of the sequence 
$q_L=\sqrt{\lambda_{\rm max}^{(L+1)} / \lambda_{\rm max}^{(L)} }$,
extracted by the Aitken delta--$2$
algorithm (exponential fit).} for the thermodynamic entropy per site
$-F(0,0)={\rm Log}\, q$, with $q=1.208..$, in very good agreement with
a previous numerical estimate \KJ, and with the exact
result \entro.

\newsec{Magnetization, critical magnetic field and the two phase hypothesis}

As mentioned above, the magnetization $M$ of the system is an
order parameter for
the flatness of the lattice. 
At $K=0, h=0$ the system is 
folded in average, with a non--vanishing folding entropy $-F(0,0)$, 
therefore the magnetization $M$ \magnet\ vanishes identically. 
On the other hand,
we will have $M \to 1$ for sufficiently large $h>0$.

%
\fig{Magnetization $M$ versus magnetic field $h$ for 
$L=4$ and $6$. The four curves correspond respectively (from
the right to the left) to
$\exp ({K/2})=1\,(K\!=\!0)\, ,1.0333..,1.0666..$ and $1.1$.
The dashed vertical lines indicate the critical magnetic field 
$h_{c}^{(L)}$ as predicted by eqn.(3.1).}{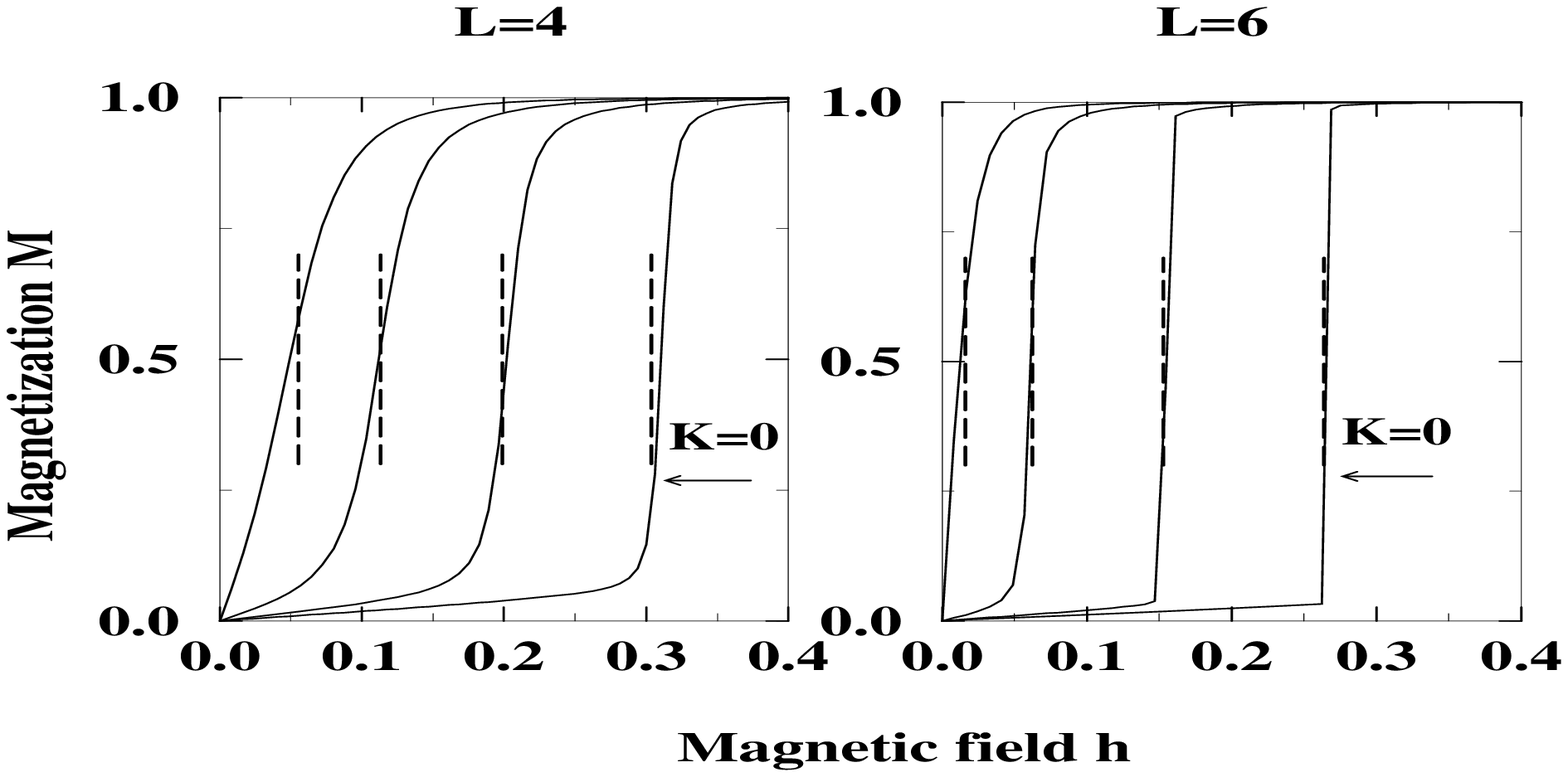}{10cm}
\figlabel\mag
%

Fig.\mag\ represents the magnetization $M$ versus
the magnetic field $h \geq 0$, computed through 
the formula \magnet\ for strips of width $L=4$ and $6$, for several
values of $K$, in the range $[0,0.2]$.
It clearly appears, and even more so for the larger $L$, that the 
magnetization tends to remain zero over a finite range of (small enough)
magnetic fields, and then abruptly jumps to $1$ when $h$ reaches
some critical value $h_{c }^{(L)}(K)$.
Moreover, this critical field $h_{c}^{(L)}(K)$ is maximal 
for $K=0$ and decreases with increasing $K$.
For a given $K$, $h_{c}^{(L)}(K)$ also
decreases with increasing $L$, eventually reaching its thermodynamic limit
$h_c(K)$ for $L \to \infty$.

This somewhat unexpectedly rapid change in $M$ is the first
tangible sign of the existence of a first order magnetic transition 
in the system.
At this point, it is reasonable to infer that in the 
thermodynamic limit $L \to \infty$, and for a given coupling $K \geq 0$,
the magnetization $M$ is exactly zero for $0 \leq h< h_c(K)$, and
exactly one for $h > h_c(K)$, being therefore discontinuous at $h=h_c(K)$,
with $h_c(K)$ a decreasing function of $K$.
Indeed all our results in the following will corroborate
this picture of a first order transition between two phases with respectively
$M=0$ and $M=1$, {\it without any possible intermediate value of 
the magnetization}.

\medskip

A first check of the above two phase hypothesis is actually
provided by the possible
{\it derivation} of the value of the critical magnetic field through
the following simple theoretical argument. 
Suppose that for large enough but finite strip width $L$, 
we can already describe the system in terms of two phases
$M=0$ and $M=1$.
In the phase $M=0$ ($h<h_{c}^{(L)}(K)$), 
the system is {\it insensitive} to the value of the magnetic
field $h$, its partition function is therefore given by 
$Z^{(L)}(K,h)\simeq Z^{(L)}(K,0)=\exp \big(-2L F^{(L)}(K,0)\big)$.
In the flat phase $M=1$ ($h>h_{c }^{(L)}(K)$), 
the partition function is that of the {\it pure state} 
with all spins up, hence reads
$Z^{(L)}(K,h)\simeq \exp \big( (3L-1)\, K + 2L \, h \big)$, 
since $3L-1$ bonds separate the $2L$ triangles.
The phase transition is then predicted to occur at the critical
value of the field $h$ where the free energies of the 
two phases are identical, namely
\eqn\crith{ h_{c}^{(L)}(K) = -F^{(L)}(K,0)
-{K \over 2} \big(3 - {1 \over L}\big) \ ,}
where
$-F^{(L)}(K,0)={1 \over 2L}{\rm Log}\, \lambda_{\rm max}^{(L)}(K,0)$
can now be calculated numerically, directly from the {\it zero magnetic
field} transfer matrix.
The corresponding predicted values of the critical field are represented on 
Fig.\mag\ by dashed vertical lines, for the various values of $K$ and $L$.
The agreement with the observed transition point
is excellent.
In the following, we shall therefore consider $h_c^{(L)}(K)$ of 
eqn.\crith\ as giving the exact location of the transition point.

\newsec{Transition line and critical coupling $K_c$}

We are now interested in understanding the thermodynamic 
critical line $h_c(K)$ separating, in the $(K,h)$ phase diagram,
the $M=0$ and $M=1$ phases.

%
\fig{The critical magnetic field $h_{c}^{(L)}(K)$ 
resulting from eqn.\crith, and the purported thermodynamic 
limit $h_c(K)$ (dashed line).}{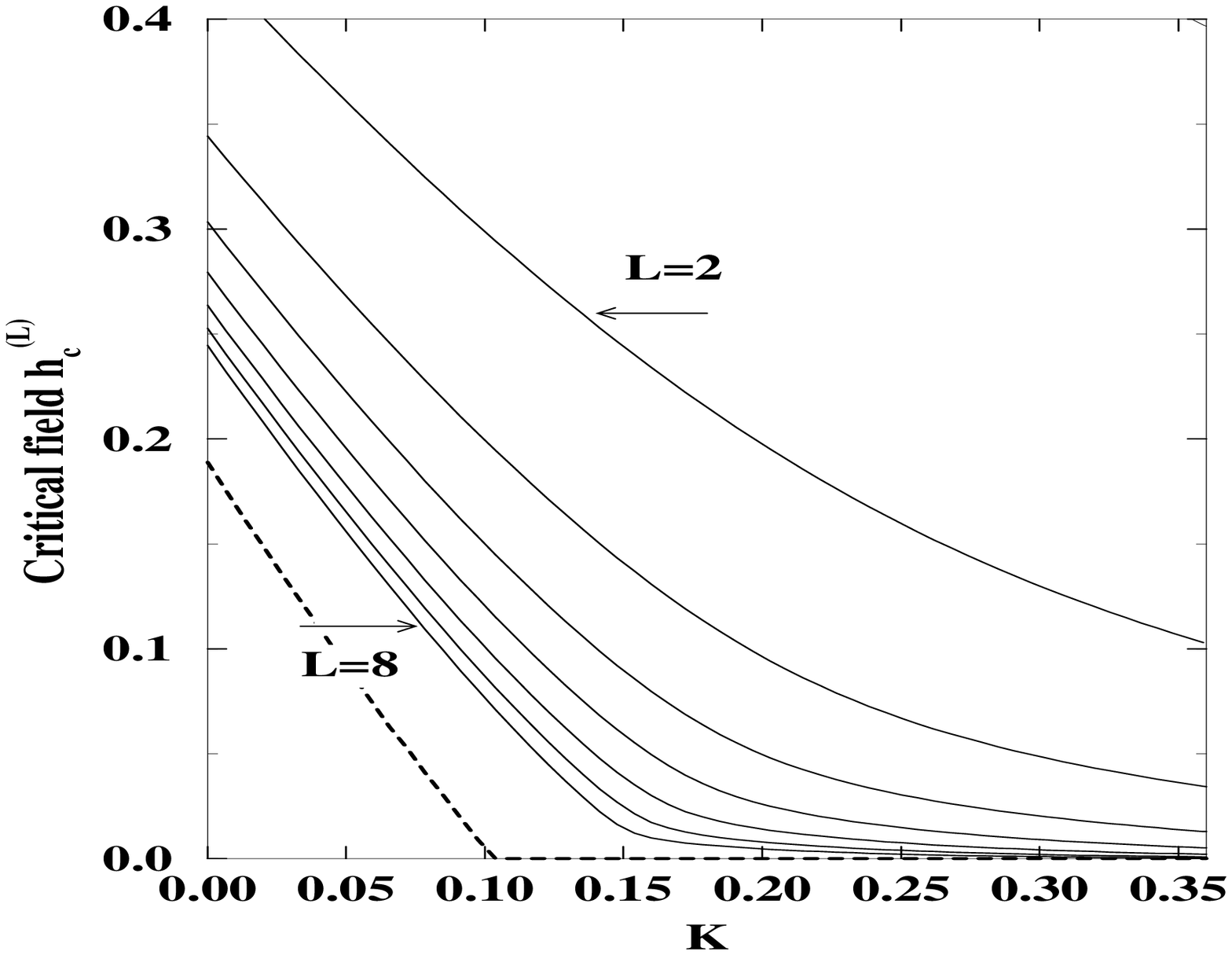}{7.cm}
\figlabel\hcrit
%

On Fig.\hcrit, we have represented the curves 
$h_{c}^{(L)}(K)$ as given by eqn.\crith\
for $L=2,3,...,8$ and $K \in [0,0.36]$.
These curves enjoy the following properties.
Eqn.\crith\ expresses the critical field $h_{c}^{(L)}(K)$ as the 
difference between the total zero field
free energy and that of a particular
state (the flat $M=1$ state): 
$h_{c}^{(L)}(K)$ is therefore positive.  
It is also clear that
$h_{c}^{(L)}(K) \to 0$ when $K \to \infty$ as the system 
gets fully ordered for strong coupling.
Finally $h_{c}^{(L)}(K)$ is a decreasing function of $K$, indeed 
\eqn\nrjL{
\eqalign{
{d h_{c}^{(L)}(K)  \over d K}&= {1 \over 2L}
\sum_{(ij)} \big( \langle \sigma_i \sigma_j \rangle_{(K,0)} -1 \big) \cr
&= E^{(L)}(K,0) -{1 \over 2}\big( 3 -{1 \over L} \big) \ , \cr}}
where the sum extends over the $3L-1$ active bonds $(ij)$ of a column 
of $2L$ triangles, and $E^{(L)}(K,0)$ is the zero field average energy per
triangle of a strip of width $L$. 
As $\langle \sigma_i \sigma_j \rangle$ is always smaller 
or equal to one, we deduce that $h_{c}^{(L)}(K)$ 
decreases with increasing $K$.

When $L \to \infty$, the curves $h_{c}^{(L)}(K)$ of Fig.\hcrit\ 
tend to the thermodynamic critical line $h_c(K)$. The properties
mentioned above for finite $L$ naturally extend to the thermodynamic
limit. The critical field, now given by
\eqn\therhcrit{h_c(K)= -F(K,0) -{3 \over 2} K \ ,}
is thus positive {\it or zero}.
At $K=0$, $h_c^{(L)}(0)$ is nothing but the entropy per triangle
$-F^{(L)}(0,0)={1 \over 2L} {\rm Log} \ \lambda_{\rm max}^{(L)}(0,0)$,
with $\lambda_{\rm max}^{(L)}(0,0)$ given in Table I. Consequently,
in the thermodynamic limit $L \to \infty$, we have the exact result
\eqn\therhcritzero{h_c(0)={\rm Log} \, q=.189...\ ,}
the transition thus taking place at a {\it non--zero} value of $h$.
By continuity, $h_c(K)$ will remain strictly positive for small positive 
$K$. Differentiating eqn.\therhcrit, we get
\eqn\nrj{
\eqalign{ 
{d h_c(K)  \over d K}
&= {3 \over 2}(\langle \sigma_i \sigma_j \rangle_{(K,0)}  -1 ) \cr
&=E(K,0) -{3 \over 2} \ , \cr}}
where $\langle \sigma_i \sigma_j \rangle$ denotes the correlation 
function of {\it any} two neighbouring spins,
and $E(K,0)$ is the zero field average energy per triangle of the system.
Again, as $\langle \sigma_i \sigma_j \rangle$ is always smaller 
or equal to one, we deduce that $h_c(K)$ is a non--increasing function.
Note that eqn.\nrj\ is nothing but the Clapeyron relation 
$$ {d \over d K} h_c(K) = -{ E_1 -E_0 \over M_1 - M_0} \ , $$
where $E_i$ (resp. $M_i$) denotes the Ising bending energy 
(resp. magnetization) in the phase $i=0$ or $1$ on each side of the
critical line.

Once $h_c(K)$ is known, the $h$ dependence of the system is determined
since
\eqn\fkh{ -F(K,h)= \left\{ \matrix{-F(K,0)& h<h_c(K)&\cr
{3 \over 2}K +h & h>h_c(K)&\cr} \right. }
In turn from eqn.\therhcrit, $h_c(K)$ is encoded in the zero field
free energy $-F(K,0)$.

%
\fig{${d h_{c}^{(L)}(K) / d K} = 
E^{(L)}(K,0)-{(3L-1)/ 2L}$, for $L=2,3,...,8$ 
and $K \in [0,0.36]$.}{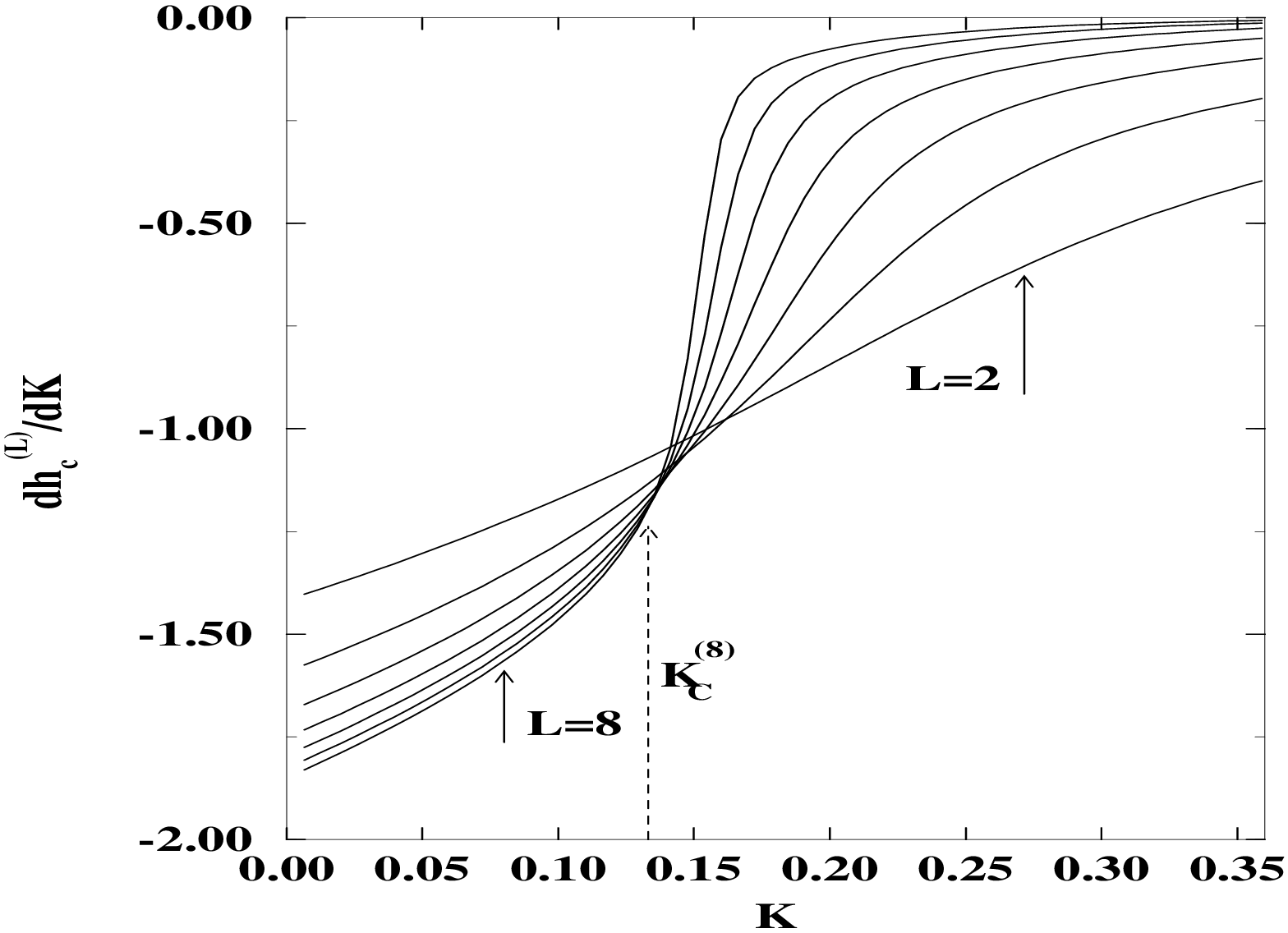}{7cm}
\figlabel\nrgy
%
A new interesting phenomenon can be read off Fig.\hcrit. 
For large enough $L$, the decrease to zero of $h_c^{(L)}(K)$ with
increasing $K$ takes place over a finite interval $[0,K_c^{(L)}]$, 
with $h_c^{(L)} \simeq 0$ for $K> K_c^{(L)}$. 
This is best seen on Fig.\nrgy\ which represents 
the curves $d h_{c}^{(L)}(K)/ d K =E^{(L)}(K,0)- {(3L-1)/ 2L}$
for $L=2,3,...,8$ and $K \in [0,0.36]$.
One clearly sees a jump in the slope of $h_c^{(L)}(K)$ from  a finite
value ($\simeq -1.2$ for $L=8$) to $0$. 
Moreover the intersections of the various curves
provide us with  estimates of the critical values $K_c^{(L)}$.
The latter decrease with $L$ and
converge to a limiting value $K_c \simeq 0.1$.

In the thermodynamic picture, this means the existence of a critical
value $K_c$ of the coupling $K$,
such that $h_c(K_c)=0$ (i.e. $-F(K_c,0)={3 \over 2} K_c$),
and thus $h_c(K)=0$ for all $K>K_c$.
On Fig.\hcrit, we have represented in dashed line the curve $h_c(K)$ as
obtained by extrapolating\foot{Like for our numerical
estimate of $q$, we obtain 
the value $h_c(K)$  
as the logarithm of the limit of the sequence 
$q_L(K)={\exp (L+1)h_c^{(L+1)}(K) / \exp Lh_c^{(L)}(K) }$,
extracted by the Aitken delta--$2$
algorithm.} to $L=\infty$ the values of $h_c^{(L)}(K)$
at fixed $K$.
This direct extrapolation confirms the emergence of a critical $K_c$
and predicts a value $K_c=0.11(1)$. From Fig.\nrgy, it corresponds
to $E(K_c,0)\simeq 0$, {\it i.e.} a vanishing nearest neighbor
spin correlation $\langle \sigma_i\sigma_j \rangle \simeq 0$.

\newsec{Phase diagram}
\nobreak
%
\fig{Phase diagram in the $(K,h)$ plane. 
Three first order lines $h=h_c(K),-h_c(K)$ ($K<K_c$) and $h=0$
($K>K_c$) separate the three phases $M=0$, $\pm 1$ and meet at
the triple point $(K_c,0)$. The dashed line
represents a constant magnetic field line, which crosses the
transition line $h=h_c(K)$ at a critical value $K_c(h)$.}{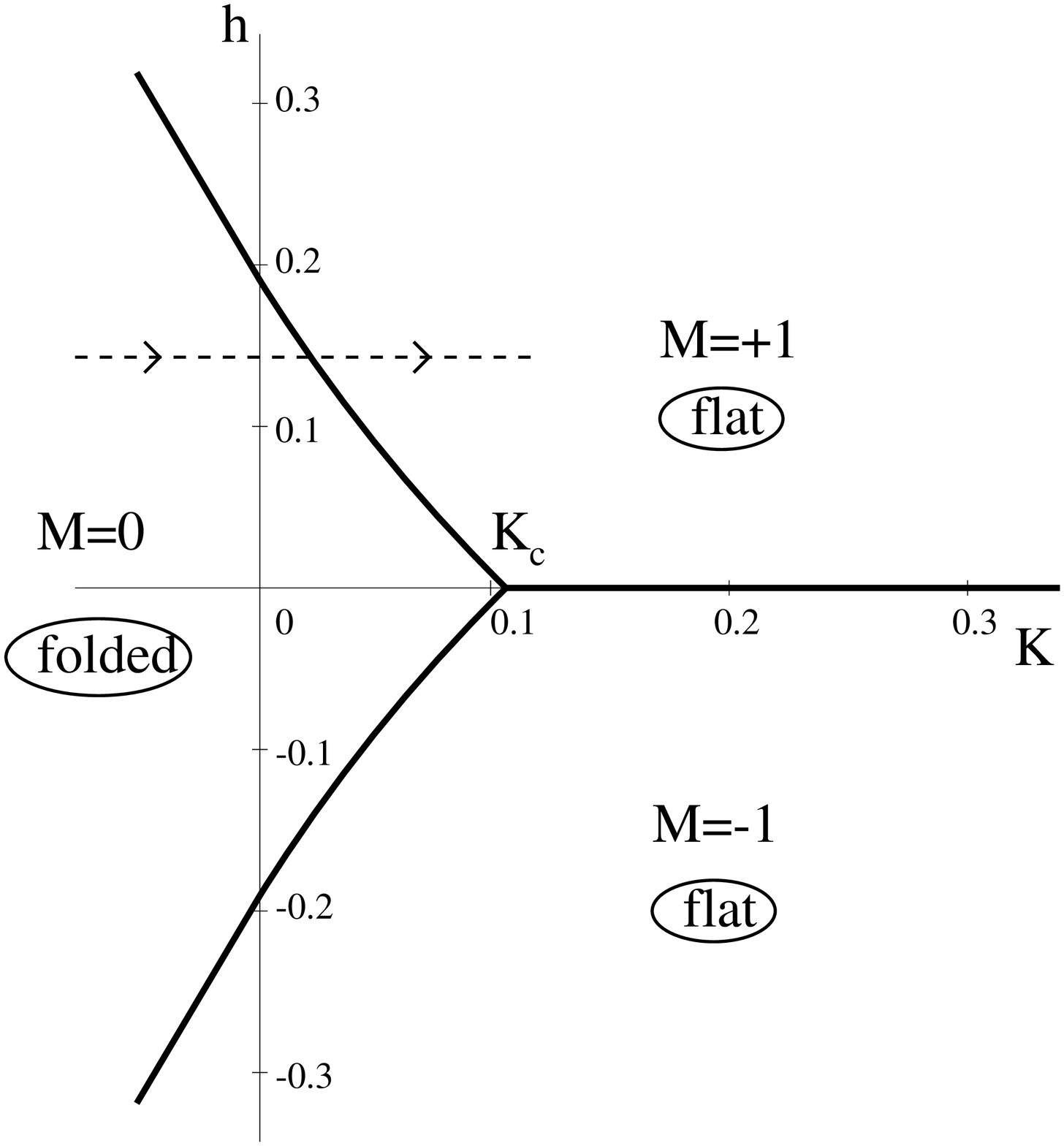}{7.cm}
\figlabel\phases
%
\nobreak
Our results are summarized on Fig.\phases, representing
the phase diagram of the sytem in the $(K,h)$ plane. 
We extended the range of $h$ and $K$ to include real negative values.
The phase diagram is clearly symmetric under $h \to -h$, while a negative
$K$ simply corresponds to an antiferromagnetic Ising coupling, which 
favorizes folding.
Three first order transition lines $h=h_c(K)$, $-h_c(K)$ ($K<K_c$),
and $h=0$ ($K>K_c$) separate the three phases $M=0,\ \pm1$.
The line $h_c(K)$ naturally extends to negative $K$ through eqn.\crith. 
For large negative $K$, the $M=0$ phase is dominated by the two pure 
antiferromagnetic states with alternating spins, with free energy
$-F(K,0) \to -{3 \over 2} K$, hence $h_c(K) \sim -3 K$ for $K \to -\infty$.
We expect nothing special to occur at $K=0$ for the line $h_c(K)$, where
we have the exact result $h_c(0)={\rm Log} \, q=.189...$. 
Instead our study predicts the existence of a triple point $(K_c,0)$  
at the {\it positive} value $K_c\simeq .11(1)$.
For the physical zero magnetic field membrane problem, this 
corresponds to a {\it first order folding transition} at $K=K_c$.  
Within the domain $M=0$, the system is insensitive to the magnetic field $h$.
Along the constant magnetic field dashed line of Fig.\phases, with $K$
increasing from $-\infty$, the free energy of
the system is $-F(K,0)$ until one reaches the critical line $K=K_c(h)$
(inverse of $h=h_c(K)$), beyond which the free energy becomes the
linear function $-F(K,h)={3 \over 2} K +h$.

%
\fig{Plot in dashed lines of 
${1 \over 2L} {\rm Log} \, \lambda_{\rm max,sub}^{(L)}(K,h)$ 
for $L=6$, $\exp(h)=1.1$ (short dashes) and $\exp(h)=1.2$ (long dashes), and
$K \in [0,0.15]$. The solid lines represent the free energies 
$-F_1^{(6)}(K,h)=h+17K/12$, and $-F_0^{(6)}(K,0)$.
The crossings occur at the critical couplings 
$K_c^{(6)}({\rm Log}\,1.2)\simeq 0.05$ and 
$K_c^{(6)}({\rm Log}\,1.1)\simeq 0.10$.}{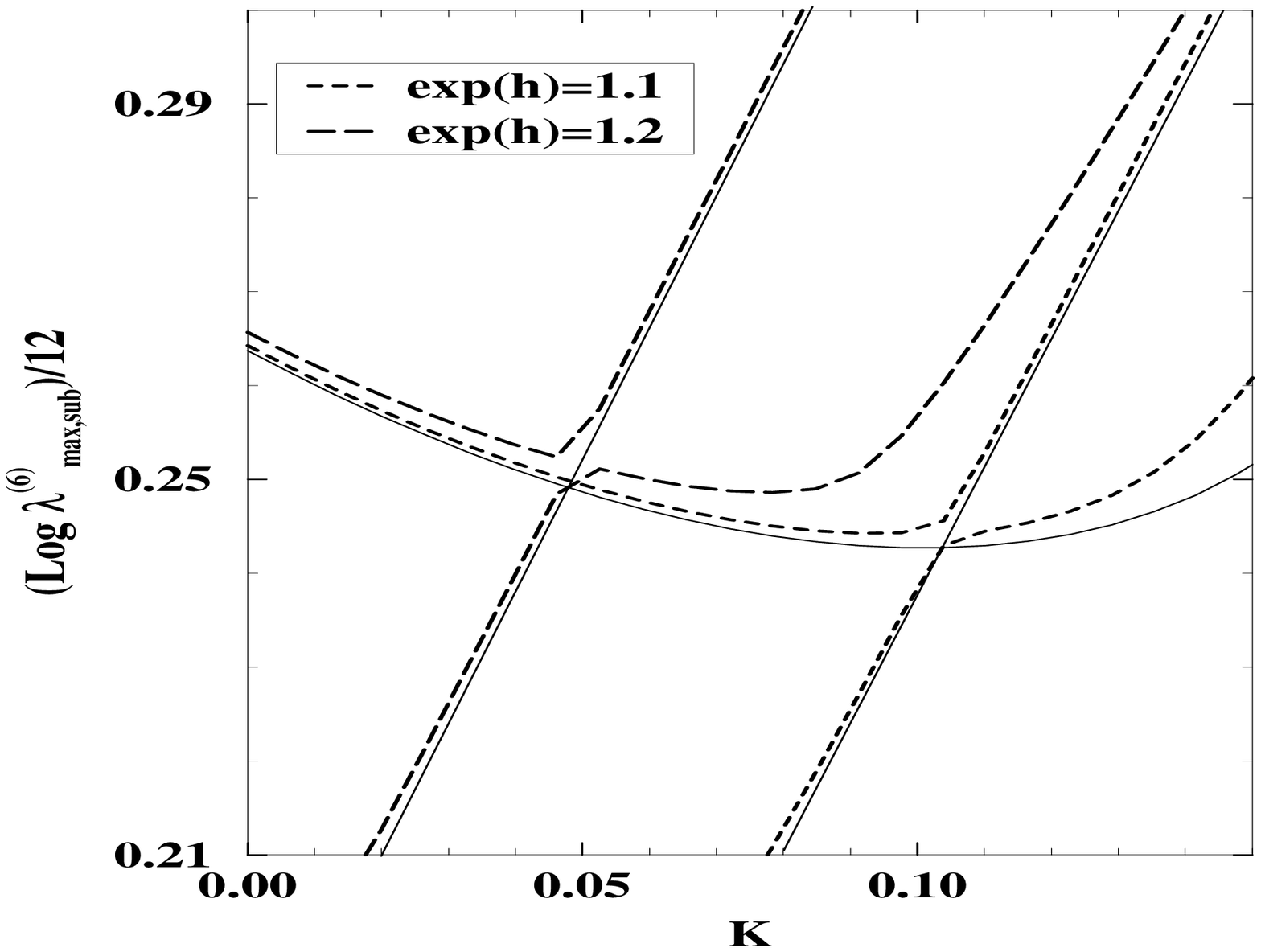}{5.5cm}
\figlabel\cross
%

As usual for first order transitions in the transfer matrix formalism, 
this change of behavior results from
the crossing of the two largest eigenvalues $\lambda_{\rm max}$ 
and $\lambda_{\rm sub}$ of the transfer matrix $T$ in
the thermodynamic limit. 
Indeed this is already visible for finite $L$, as exemplified on Fig.\cross,
where we plot for $L=6$ (in dashed lines)
the two leading eigenvalues of $T^{(6)}(K,h)$
for two different positive values of $h$, in the vicinity of the 
corresponding critical
points $K_c^{(6)}(h)$.
One clearly sees the exchange of the two eigenvalues with 
$$
{1 \over 2L} {\rm Log} \left\{ \matrix{ \lambda_{\rm max} \cr
\lambda_{\rm sub} \cr} \right\}=\left\{ \matrix{\left\{
\matrix{-F_0^{(L)} \cr
-F_1^{(L)} \cr} \right\}
& K<K_c^{(L)}(h) & \cr
\left\{\matrix{ -F_1^{(L)} \cr
-F_0^{(L)} \cr} \right\}
& K>K_c^{(L)}(h) &\cr} \right. \ , $$
where $-F_1^{(L)}(K,h)=h+(3L-1)K/2L$, and $-F_0^{(L)}(K,h)=-F^{(L)}(K,0)$
denote respectively the free energy in the 
phase with magnetization $M=1$ and $0$,
represented for $L=6$ on Fig.\cross\ in solid lines.

%
\fig{Finite (dashed) and zero (solid) field energy versus Ising coupling, 
for $L=6$ and $\exp(h)=1.2,1.1,1$ respectively, and 
$K \in [0,0.36]$.}{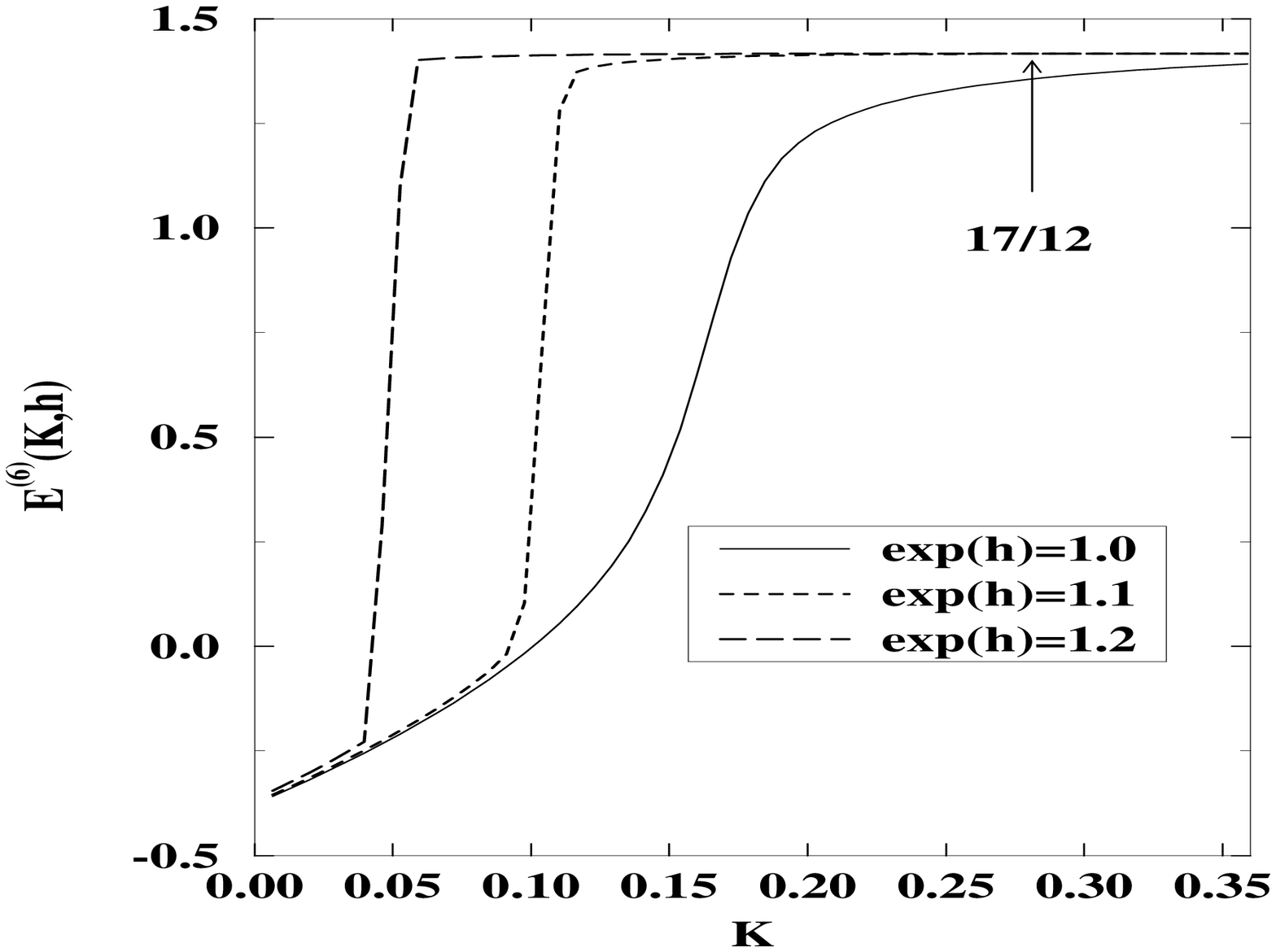}{6.cm}
\figlabel\energum
%

The change of behavior of the free energy along a 
constant--field line at $K_c(h)$
is confirmed at finite $L$ by the plot of its 
derivative w.r.t. $K$,
the Ising energy $E^{(L)}(K,h)=-\partial_K F^{(L)}(K,h)$, which we 
represent on Fig.\energum\ for $\exp(h)=1,1.1,1.2$ and $L=6$.
We clearly see that the finite field energies $E^{(6)}(K,h)$
exactly match
the zero field energy $E^{(6)}(K,0)$ before $K$ reaches the critical value
$K_c^{(6)}(h)$, where they abruptly jump (the more so for higher $h$)
to the asymptotic value $(3\times 6-1)/(2\times 6)=17/12$.

%
\fig{Specific heat $C_{\rm v}^{(6)}(K,h)$ for $\exp(h)=1$
(thick solid line), $1.1$ and $1.2$ (dashed lines) 
and $C_{\rm v}^{(L)}(K,0)$ for $L=2,3,...,8$, for $K \in [0,0.36]$.}{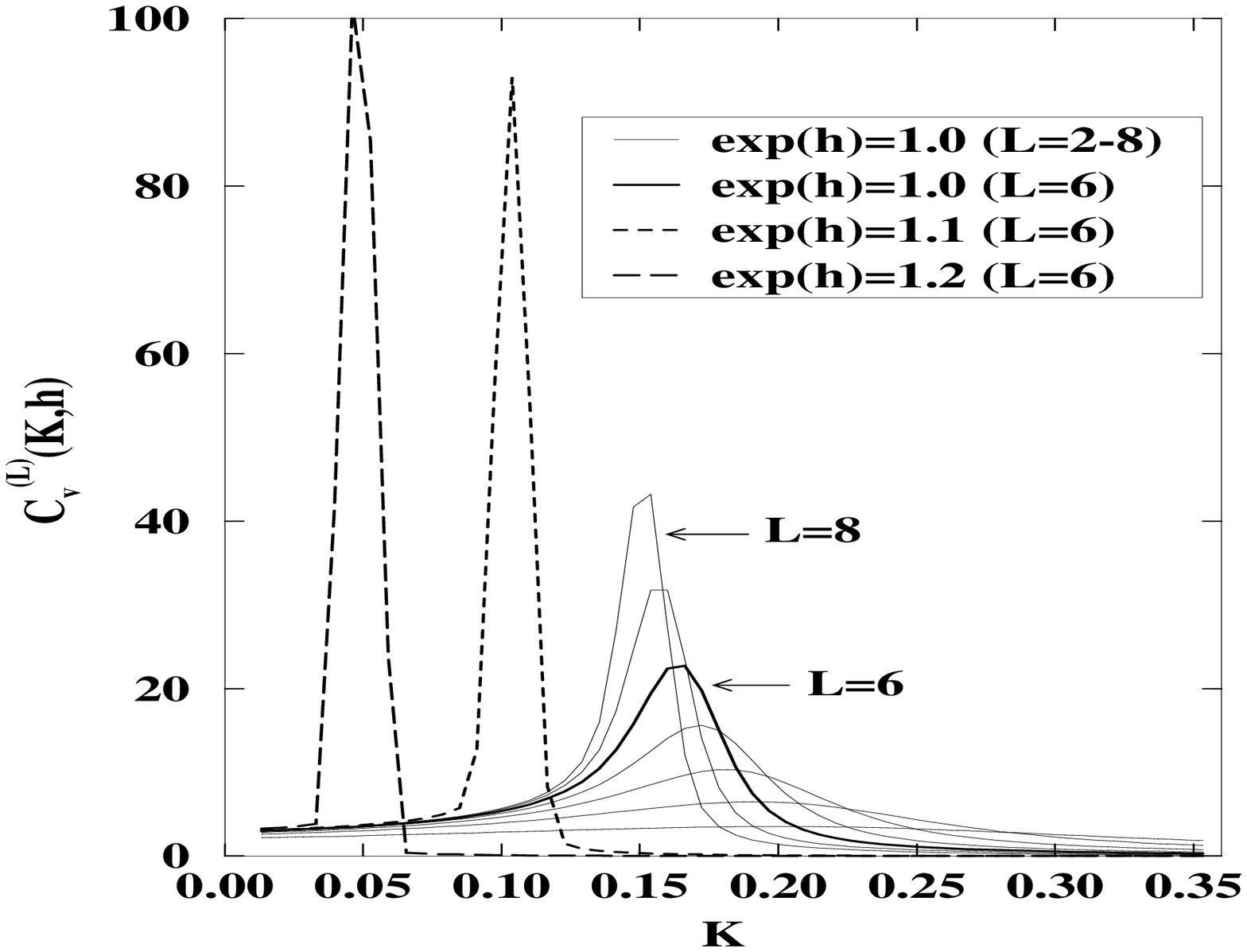}{7.cm}
\figlabel\spec
%

The corresponding specific heat 
$C_{\rm v}^{(L)}(K,h)=\partial_K E^{(L)}(K,h)$
is represented on Fig.\spec\ for $L=6$ in dashed ($\exp(h)=1.1,1.2$)
and thick solid ($\exp(h)=1$) lines. As expected, the finite field 
specific heats match the zero field specific heat until $K$ reaches $K_c^{(6)}(h)$,
where they exhibit a delta--function peak, before immediately (the more
so for higher $h$) reaching a zero value.  
The peak for $h=0$ seems to be
qualitatively different from that for finite $h$.
To see how this peak develops as the size $L$ grows, we have
represented on the same figure the zero field specific heat 
for $L=2,3,...,8$ in thin solid lines. 
The smoothness of the curves
might be the sign of some extra divergence on top of the 
delta--function in the thermodynamic specific heat $C_{\rm v}(K,0)$,
possibly of the form $C_{\rm v}(K,0) \sim (K_c -K)^{-\alpha}$ when $K$ 
approaches the triple point from below.
Our data do not allow us at present to reach any conclusion on this point.

\newsec{Discussion}

The main result of this paper is the phase diagram of Fig.\phases. 
In particular, the existence of a triple point $(K=K_c>0,h=0)$ at
the boundary of the three phases with magnetization $M=0$, $\pm 1$
is a non--trivial outcome of our study. 
It is interesting to note that a very similar
picture can be obtained 
{\it exactly}\foot{An exact solution can be obtained
for instance by using a transfer matrix which is diagonal in terms
of folded line variables.} 
in the case of the folding of the {\it square} lattice.  
As already noted in \KJ, the thermodynamic entropy of folding
$-F_{\rm square}(0,0)$
of the square lattice {\it vanishes}, due to the very strong constraint
that, for this particular lattice, folds must propagate
along straight lines, all the way through the lattice. A folded
state of the square lattice is entirely specified by the data of its folded
horizontal and vertical lines.
For a square lattice of size $L \times L$, 
this leads to a free energy per square
$-F_{\rm square}(0,0)=\lim_{L \to \infty}{1 \over L^2} \hbox{Log}\ 4^L =0$,
as annouced. 
Like in the triangular case, the square lattice folding problem 
is easily transformed into a (face) square lattice Ising spin system, 
with the local constraint that there are exactly
$0$, $2$ or $4$ spins up around each vertex.
The zero field thermodynamic free energy $-F_{\rm square}(K,0)$ per square
at an arbitrary value of the reduced Ising coupling $K$ is easily obtained
as follows:
at $K=0$, it vanishes; 
for $K \to \infty$, it tends to the flat state value $2 K$;
for $K \to -\infty$, it tends to the completely folded state
value $-2K$. {}From the usual convexity property of $-F$, 
we conclude that necessarily
$$ -F_{\rm square}(K,0)= 2 |K| \ .$$
The free energy appears here simply as a competition between the contribution
of the completely folded state $-F_0(K,0)=-2K$ and that of the flat state
$-F_1(K,0)=2 K$, with $-F_{\rm square}(K,0)={\rm Max} (-F_0,-F_1)$.
Analogously, in the presence of a finite positive magnetic field $h$,
the free energy results from the competition between the contribution of the 
completely folded state, insensitive to $h$, $-F_0(K,h)=-2K$, and that of 
the flat (all spins up) state $-F_1(K,h)=2K+h$. This yields
$$-F_{\rm square}(K,h)={\rm Max}\big(-F_0(K,h),-F_1(K,h)\big)=
\left\{ \matrix{-2K & K<-{h / 4} & \cr
2K+h & K>-{h / 4} & \cr} \right. \ .$$
Consequently the system undergoes a first order phase transition
from the completely folded state with magnetization 
$M=0$ to the flat state with $M=1$, along the critical line
$$h_{c,{\rm square}}(K)=2(|K|-K) \ ,$$
the square lattice analogue of eqn.\crith.

%
\fig{Phase diagram of the square lattice folding problem.}{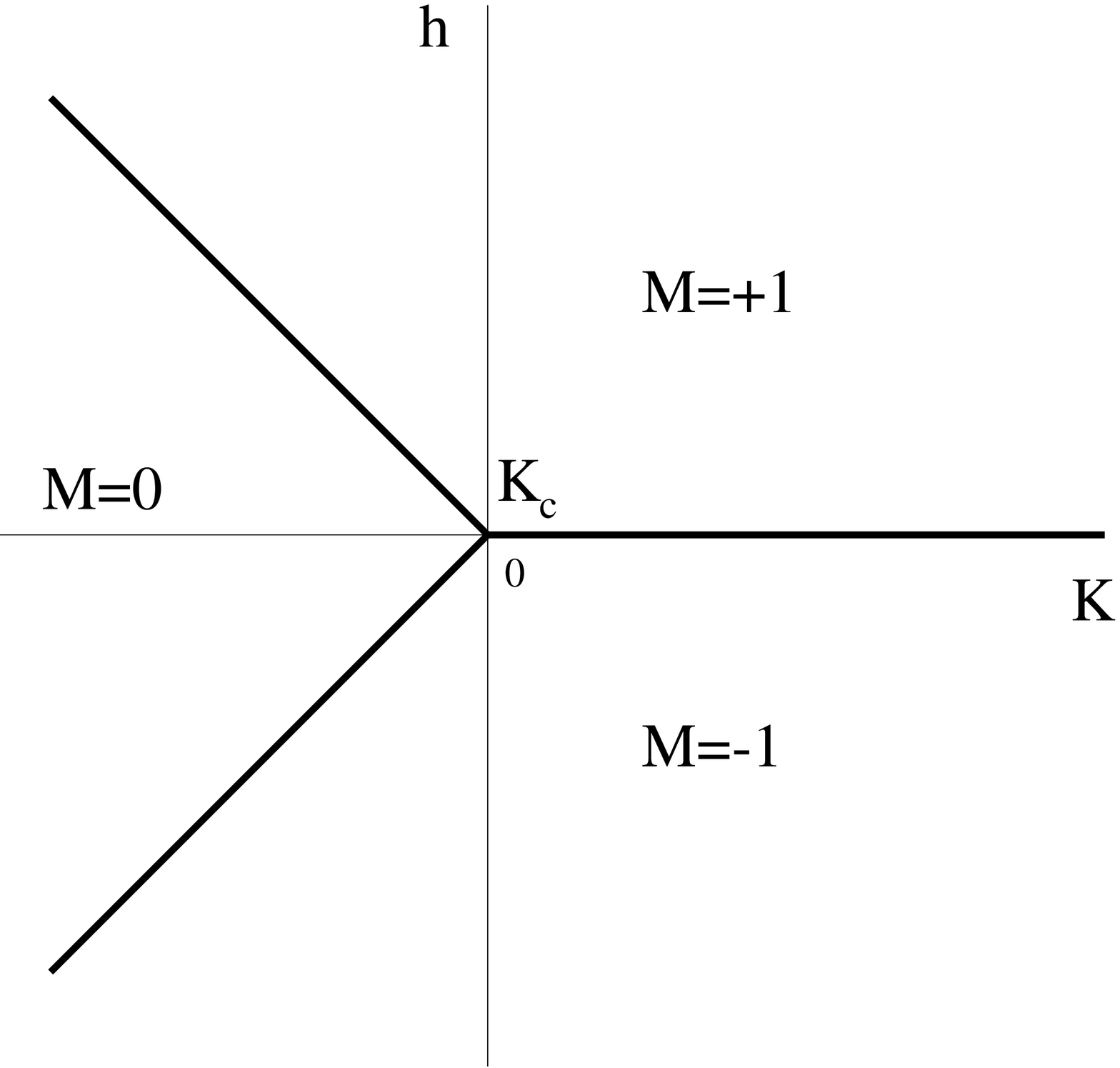}{6.cm}
\figlabel\phasq
%

As shown on the resulting phase diagram of Fig.\phasq, 
the qualitative behavior of the system is very similar to the triangular case
discussed above, corroborating a posteriori the diagram of Fig.\phases.
A crucial difference is that the triple point now sits at the origin of the $(K,h)$ plane, i.e. $K_{c,{\rm square}}=0$.
Consequently no folding transition occurs at positive $K$
for the square lattice, which remains flat.
That $K_c$ is positive for the 
triangular lattice follows directly from the positivity of the 
entropy at $K=0$, which makes the triangular case more interesting.
 
A surprising outcome of our study of the triangular lattice case
is the absence of intermediate magnetization states between $M=0$ and $M=1$. Denoting by $-F_m(K,0)$ the contribution
of the configurations with magnetization $M=m$ to the zero field free energy, 
a sufficient condition for having no intermediate magnetization
is that the critical field $h_{c,m}(K)$
governing the transition between the $M=0$ and an hypothetical $M=m$
phase be always larger than $h_{c,1}(K)=h_c(K)$, for $K<K_c$, namely
$$h_{c,m}(K)={ -F(K,0)+F_m(K,0) \over m} > h_{c,1}(K) 
= -F(K,0)+{3 \over 2}K \ ,$$
with $m<1$.
The freezing of the system for $K>K_c$ in, say, the $M=1$ phase is probably
a consequence of the intrinsically non--local character of the folding
constraint, preventing the creation of bounded domains of down spins
inside a groundstate of up spins.

The stability of the $M=0$ phase at positive $K$ 
in the presence of a magnetic field 
$h<h_c(K)$ is more surprising.
A way of refining the study of the $M=0$ phase is to introduce a new
order parameter, the staggered magnetization 
$$M_{\rm st}={1 \over N_t}\langle \big( \sum_{\bigtriangleup} \sigma_i -
\sum_{\bigtriangledown} \sigma_i \big) \rangle ,$$
where the sum alternates between triangles
pointing up and down in the lattice.
One can then distinguish between the disordered folded state $M_{\rm st}=0$
and a compactly ordered folded state $M_{\rm st}>0$ where the triangles
start to pile up. 
In the presence of a staggered magnetic field, we expect the system to
behave qualitatively like an antiferromagnetic (face) 
triangular Ising model, with a continuous ``piling"
transition at some negative value of the coupling $K_{c,{\rm st}}<0$.

%
\fig{Heat capacity $C_{\rm v}^{(L)}(K,0)$ for $L=4,6,8$ and 
$K \in [-0.8,0]$.}{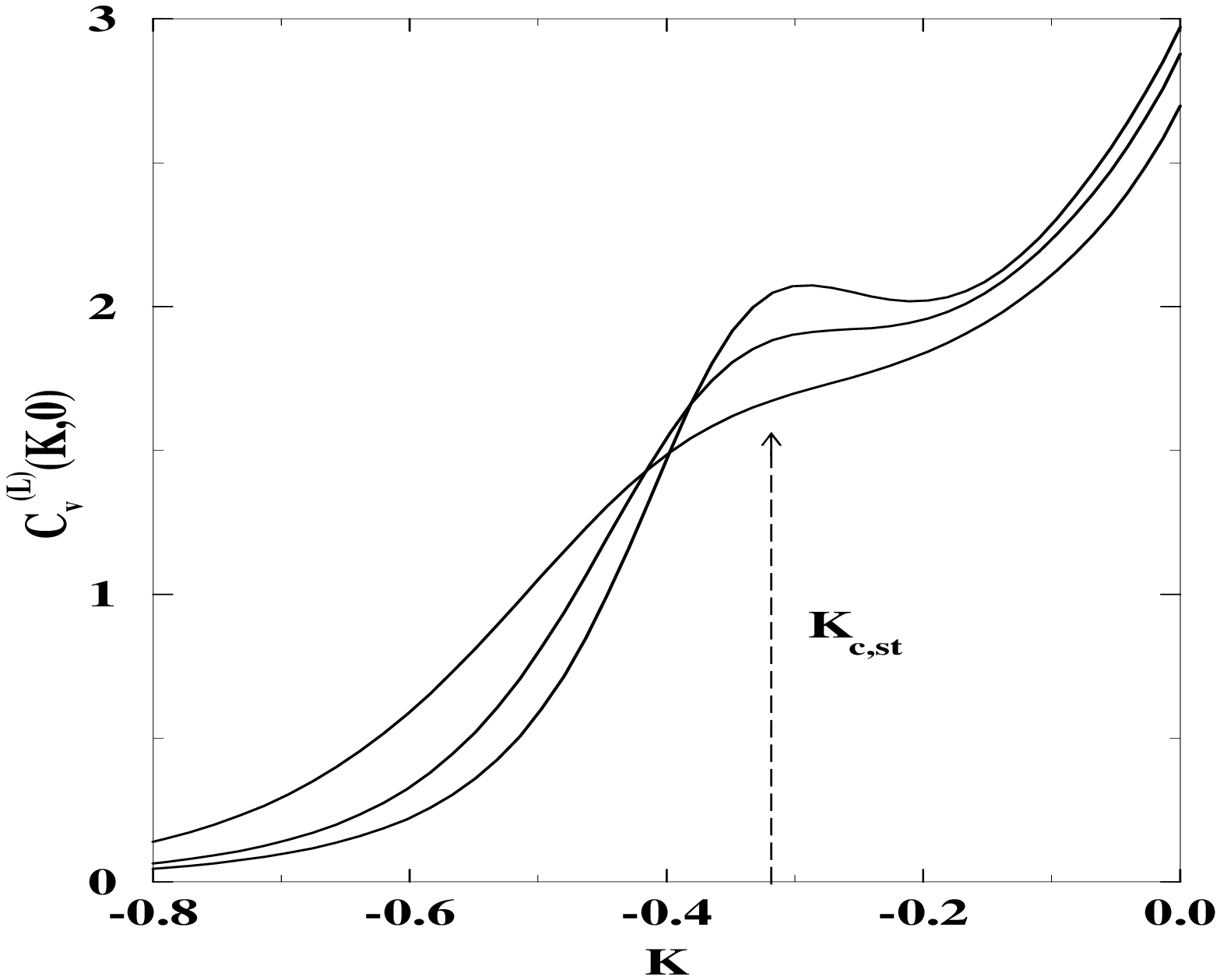}{7.5cm}
\figlabel\af
%

\noindent Indeed, as explained in \DIG, the situation is very different when we
start to {\it unfold} the antiferromagnetic $M_{\rm st}=1$ 
groundstate than when we try to
{\it fold} the ferromagnetic $M=1$ one.
Local deformations of the completely folded 
groundstate are allowed and enable a ``low temperature" (large negative $K$)
expansion in terms of a gas of loops of unfolded bonds, quite similar
to the standard loop gas expansion of the Ising model.
As far as numerical results are concerned, we only see on Fig.\af\ 
the slow emergence
of a peak in the specific heat at a value $K_{c,{\rm st}}\simeq-0.3$.
At $K=0$, an exact solution \BAX\ predicts a disordered folded state
$M_{\rm st}=0$ with finite staggered susceptibility.

\newsec{Conclusion}

In this paper, we have derived the phase diagram of the constrained spin
system describing the folding of the triangular lattice. 
In the presence of a magnetic field, we found 
a critical line along which a first order transition takes
place between a zero magnetization phase and the $M=1$ pure state, 
terminating at a triple point $(K=K_c,h=0)$. 
This transition persists at zero magnetic field, now driven by the coupling $K$,
and can be interpreted as a {\it first order folding transition}
between a folded phase and the completely flat state of the lattice.
The latter is reminiscent of the crumpling transition of tethered membranes 
[\xref\NP -\xref\DG ], which is however continuous rather than first order.

The phase diagrams of Fig.\phases\ and \phasq\ are very far from the
usual unconstrained Ising ones. It would be interesting to investigate
the role of the local folding constraint by applying it gradually to 
the Ising model, and by looking at the deformation of the phase diagram.
In this framework, one should be able to follow the evolution of the system 
from a continuous second order phase transition to a first order one.
Note that in the square case both the Ising and constrained models are
special cases of the eight vertex model in an electric field, yet to be
solved.

\vskip 1.cm
\leftline{\bf Acknowledgements}

We are grateful to A. Billoire, R. Lacaze and V. Pasquier for
useful discussions, and especially to O. Golinelli for 
sharing with us his experience in numerical calculations.
We thank Th. Jolic\oe ur for a careful reading of the manuscript.

\listrefs

\end